\documentclass{elsart3p}

\usepackage{epsfig}

\usepackage{amssymb,amsmath}
\usepackage{multicol}
\newcommand{\Spin}{{\mathrm{Spin}}}

\usepackage{slashbox}
\usepackage[usenames,dvipsnames]{color}

\newcommand{\hotimes}{{\hat{\otimes}}}

\newcommand{\Tr}{\mathrm{Tr}}
\newcommand\calA{\mathcal{A}}

\newcommand\calH{\mathcal{H}}
\newcommand\calK{\mathcal{K}}
\newcommand\calL{\mathcal{L}}

\newcommand\calS{\mathcal{S}}

\newcommand\bbC{\mathbb{C}}

\newcommand{\sixj}[4]{\left( \begin{array}{@{}c@{\;}c@{}}
#1 & #2 \\
#3 & #4
\end{array}\right)}

\begin{document}

\begin{frontmatter}



\title{Space and time dimensions of algebras with
applications to Lorentzian noncommutative geometry
and quantum electrodynamics}

\author[IMPMC]{Nadir Bizi},
\ead{nadir.bizi@impmc.upmc.fr} 
\author[IMPMC]{Christian Brouder},
\author[EPF]{Fabien Besnard},
\address[IMPMC]{Sorbonne Universit\'es,
   UPMC Univ Paris 06,
   UMR CNRS 7590,
Mus\'eum National d'Histoire Naturelle, IRD UMR 206,
  Institut de Min\'eralogie, de Physique des Mat\'eriaux et de
 Cosmochimie, 
 4 place Jussieu, F-75005 Paris, France.
}%
\address[EPF]{P{\^o}le de recherche M.L. Paris,
 EPF, 3~bis rue Lakanal,
 F-92330 Sceaux, France.
}%

\begin{abstract}
An analogy with real Clifford algebras on even-dimensional
vector spaces suggests to assign 
a couple of space and time dimensions
modulo 8 to any algebra (represented
over a complex Hilbert space) containing two self-adjoint
involutions and an anti-unitary operator with specific
commutation relations.

It is shown that this assignment is compatible with
the tensor product: the space and time dimensions
of the tensor product are the sums of the space and time
dimensions of its factors.
This could provide
an interpretation of the presence of such algebras
in $PT$-symmetric Hamiltonians or the description
of topological matter.

This construction is used to build an indefinite
(i.e. pseudo-Riemannian)
version of the spectral triples of noncommutative geometry,
defined over Krein spaces instead of Hilbert spaces.
Within this framework, we can express the
Lagrangian (both bosonic and fermionic) of a
Lorentzian almost-commutative spectral triple.
We exhibit a space of physical states that solves
the fermion-doubling problem.
The example of quantum electrodynamics is described.
\end{abstract}
\today
\end{frontmatter}

\date{\today}
 
%%%%%%%%%%%%%%%%%%%%%%%%%%%%%%%%%%%%%%%%%%%%%%%%%%%%%%%%%%%%%%%%%%%%%

\section{Introduction}
Clifford algebras are at the heart of the description
of matter not only because fermions (spinors) are their irreducible 
representations, but also because they classify 
topological insulators and
superconductors~\cite{Morimoto-13,Chiu-16}.
They are also used as a template
for deeper structures, such as
$K$-theory~\cite{Atiyah-64,Karoubi}
or noncommutative geometry~\cite{Connes-95-reality},
that pervade physics from topological matter
to disordered systems and the
standard model of particles.

The main aim of this paper is to describe a 
pseudo-Riemannian analogue of noncommutative 
geometry, but on the way we put forward
a procedure to assign a space dimension and a time dimension
to a class of algebras. What we need is
\begin{itemize}
\item A complex Hilbert space $\calH$. 
\item A self-adjoint involution $\chi$ 
(i.e. $\chi^2=1$)
defining the parity of operators:
 an operator $a$ on $\calH$ is 
even if $\chi a \chi=a$ and odd
if $\chi a \chi=-a$. For example,
$\chi$ can be the chirality operator or
the inversion symmetry.
\item A second self-adjoint involution $\eta$,
which can be the flat-band Hamiltonian
$\mathrm{sign}H$~\cite{Bernevig} or a fundamental symmetry.
\item An  anti-linear (charge conjugation) map $J$ 
such that $J^\dagger J=1$ and $J^2=\epsilon=\pm 1$.
\item Specific commutation or anticommutation relations between
$\chi$, $\eta$ and $J$ 
  defined by three signs
   $(\epsilon'',\kappa,\kappa'')$ introduced in
  Eqs.~\eqref{Jchi} to \eqref{etachi}.
\end{itemize}
In a Clifford algebra $C\ell(p,q)$ such that
$p+q$ is even, the
four signs $(\epsilon,\epsilon'',\kappa,\kappa'')$ 
determine a pair of space and time dimensions $(s,t)$ modulo 8
in a unique way.  
We propose to assign the same dimensions $(s,t)$
to any algebra satisfying the same relations
between $\chi$, $\eta$ and $J$. 
This 
is similar to the way Atiyah related the 
$KO$-dimension to $p-q\mod 8$ in Clifford
algebras~\cite{Atiyah-64}.
These two dimensions solve the question
whether indefinite spectral triples have a notion
corresponding to the $KO$-dimension~\cite{Dungen-15}. 

Such an assignment is meaningful because it is compatible
with the tensor product: the dimensions corresponding
to the signs of the (graded) tensor product $A_1\hat\otimes A_2$
of two such algebras are the sum of the dimensions
of $A_1$ and $A_2\mod 8$.

When we apply this to a spectral triple of noncommutative
geometry, $\chi$ is the usual chirality operator,
$\eta$ is a fundamental symmetry defining a Krein-space
structure and $J$ is the usual charge conjugation.
Our approach allows us to assign a space-time dimension
to the finite algebra of the almost-commutative
spectral triple of models of particles.

The paper starts with a description of 
$\chi$, $\eta$ and $J$ in a Clifford algebra,
which sets up the correspondence between
commutation relations and space-time dimensions.
Then, this correspondence is shown to hold for
more general algebras by proving that it is compatible
with the graded tensor product of algebras.
In section 4, we introduce Krein spaces, which are
the natural generalizations of Hilbert spaces
associated to spinors on pseudo-Riemannian manifolds.
Section 5 defines the corresponding generalized
spectral triples, that we call \emph{indefinite spectral triples}.
This framework is then applied to define the
spectral triple of Lorentzian quantum electrodynamics (QED)
and its Lagrangian.

\section{Automorphisms of Clifford algebras}
We investigate the commutation relations of
three operators in Clifford algebras over
vector spaces of even dimension $2\ell$.
They are simple algebras whose irreducible 
representation is $S\simeq\bbC^{2^\ell}$.

The Clifford algebra $C\ell(p,q)$ with
$p+q=2\ell$ is the algebra generated by $2\ell$
matrices $\gamma_j$ over $S$ such that
$\gamma_i\gamma_j+\gamma_j\gamma_i=2\epsilon_j\delta_{ij}$, 
where $p$ coefficients $\epsilon_j$ are equal to +1
and $q$ are equal to -1.

An \emph{indefinite inner product} on a complex
vector space $V$ is a non-degenerate sesquilinear form on $V$ which
satisfies $(v,u)=\overline{(u,v)}$. It is indefinite
because we do not assume that $(v,v)>0$ if $v\not=0$.
It was shown~\cite{Baum-81,Robinson-88,Harvey-90} 
that $S$ can be equipped with
two indefinite inner products $(\cdot,\cdot)_+$ and
$(\cdot,\cdot)_-$ such that 
\begin{eqnarray}
(\gamma_j \phi,\psi)_\pm &=& \pm (\phi,\gamma_j \psi)_\pm,
\label{Robinsonpm}
\end{eqnarray}
for every $\gamma_j$, and every elements
$\phi$ and $\psi$ of 
$S$.
These two inner products are unique up to
multiplication by a real factor and 
they are invariant under the action of $\Spin(p,q)^+$,
the connected component of the identity in $\Spin(p,q)$.
Moreover, $S$ can be equipped with two
charge conjugations $J_\pm$, which are
anti-linear maps such that
$J_\pm \gamma_j=\pm\gamma_j J_\pm$.

To give a concrete representation of the
indefinite inner products and the charge conjugations,
we equip the complex vector space $S$ with its standard 
(positive definite) scalar product $\langle \cdot,\cdot\rangle$.
By using results scattered in the
literature~\cite{Baum-81,Pauli-36,Kugo-83,Wetterich-83,Andrade-01,Shirokov-13},
we can state the following.
In all representations used in practice, the
gamma matrices satisfy
$\gamma_j^\dagger=\epsilon_j \gamma_j=\gamma_j^{-1}$,
where $\dagger$ denotes the adjoint with respect
to the scalar product, 
and $\overline{\gamma_j}=\zeta_j \gamma_j$,
where the overline denotes complex conjugation
and $\zeta_j$ is $+1$ ($\gamma_j$ is real) or $-1$ 
($\gamma_j$ is imaginary).
Shirokov showed~\cite{Shirokov-13} that these
representations can be classified by
the number of their symmetric gamma matrices
(i.e. such that $\gamma_j^T=\gamma_j$) modulo 4, which
is equal to $\ell+\zeta\mod 4$, where $\zeta$ can take
the value 0 or 1.
Both cases are useful in practice. For
example, the most common representations
of $C\ell(1,3)$ are the Dirac, Majorana
and chiral gamma matrices~\cite{Itzykson}.
The number of symmetric matrices of these
representations is 2,3 and 2, corresponding
to $\zeta=0,1$ and 0.

Then, it can be shown that the chirality matrix
$\chi=i^{(p-q)/2} \gamma_1\dots\gamma_{2\ell}$
anticommutes with all $\gamma_j$ and satisfies 
$\chi^\dagger=\chi$, $\chi^2=1$ and
$\overline{\chi}=(-1)^\zeta \chi$.
To construct the indefinite inner products
we define linear operators 
$\eta_\pm$ (called
\emph{fundamental symmetries}) such that
$(\phi,\psi)_\pm=\langle \phi,\eta_\pm\psi\rangle$,
$\eta_\pm^\dagger=\eta_\pm$ and $\eta_\pm^2=1$.
From Eq.~\eqref{Robinsonpm}:
$\eta_\pm \gamma_j^\dagger\eta_\pm=\pm \gamma_j$.
These fundamental symmetries are unique up 
to a sign and are built as follows.
Let $M_s$ be the product of all self-adjoint gamma matrices
and $M_{as}$ the product of the anti-self-adjoint ones, then
\begin{eqnarray*}
\eta_+ &=& i^{q(q-1)/2} \chi^q M_{as},
\\
\eta_- &=& i^{p(p+1)/2} \chi^{p} M_{s}.
\end{eqnarray*}

To construct the charge conjugations $J_\pm$,
let $n_i$ be the number of imaginary gamma
matrices and $N$ their product.
Then
\begin{eqnarray*}
J_+ &=& i^{(1-\zeta)(p-q)/2} \chi^{n_i} N K,
\\
J_- &=& i^{(1-\zeta)(p-q)/2+\zeta} \chi^{n_i+1} N K,
\end{eqnarray*}
where $K$ stands for complex conjugation.
The relation 
$\zeta=n_i+(p-q)/2\mod 2$ holds. The ``charge conjugations''
$J_\pm$ satisfy $J_\pm \gamma_j=\pm \gamma_j J_\pm$,
$J_\pm^\dagger J_\pm=1$ and $J_\pm$ commutes with
complex conjugation.  These conditions determine
$J_+$ and $J_-$ uniquely up to a sign.

In every Clifford algebra $C\ell(p,q)$ with
$p+q$ even, we can choose $J=J_+$ or $J=J_-$
and $\eta=\eta_+$ or $\eta=\eta_-$. This makes
four possible conventions for which we have:
\begin{eqnarray}
J^2 &=& \epsilon,\label{Jdeux}\\
J\chi &=& \epsilon''\chi J,\label{Jchi}\\
J\eta &=& \epsilon\kappa \eta J,\label{Jeta}\\
\eta \chi &=& \epsilon''\kappa''\chi\eta,
  \label{etachi}
\end{eqnarray}
where
$\epsilon=(-1)^{n(n+2)/8}$,
$\epsilon''=(-1)^{n/2}$,
$\kappa=(-1)^{m(m+2)/8}$ and
$\kappa''=(-1)^{m/2}$.
Note that $\epsilon$ and $\epsilon''$ are the
same functions of $n$ as $\kappa$ and 
$\kappa''$ are of $m$.
These signs were defined so that
$\epsilon$ and $\epsilon''$ agree with
Connes' $KO$-dimension 
tables~\cite{Connes-95-reality,Connes-Marcolli}.
Related tables can be found in the
literature~\cite{Varlamov-15,Budinich-16,Schulz-16}.

The values of the $KO$ dimension $n$ and of
the new dimension $m$ in terms
of the conventions are given in Table~\ref{tabconv}.
Note that $m$ and $n$ are defined modulo 8.

In physics the Dirac operator is written
$D=i\gamma^\mu \nabla_\mu$ in the so-called
West-coast convention~\cite{Duff-06}
and $D=\gamma^\mu \nabla_\mu$ in the East-coast 
one.
Charge-conjugation symmetry requires $JD=DJ$ and 
the reality of the fermionic Lagrangian implies
that $D$ is self-adjoint with respect to the indefinite
inner product. Thus, the West-coast convention
corresponds to $J=J_-$ and $\eta=\eta_+$ while the
East-coast one to $J=J_+$ and $\eta=\eta_-$~\cite{Berg-01}.
This is related to the signature of the metric.
Indeed, in Minkowski space time we want 
plane wave solutions of the (massive) Dirac
equation: $\psi(x)=ue^{i k_\mu x^\mu}$. 
Compatibility with the dispersion relation
$k^2=m^2_e$, where $m_e$ is the fermion mass, 
implies the metric $(+,-,-,-)$ for the West-coast
convention and $(-,+,+,+)$ for the East-coast 
one~\cite{Berg-01}.

In Euclidean space, we can be interested in 
the real solutions $\psi(x)=ue^{k_\mu x^\mu}$.
The metric $(-,-,-,-)$ corresponding to $J=J_-$ 
and $\eta=\eta_-$ is often used and we call
it the \emph{North-coast convention} because
Euclid lived on the coast of North Africa.
The remaining possibility is $J=J_+$
and $\eta=\eta_+$ that we call the South-coast
convention.

In Lorentzian spacetime we need plane waves
to describe scattering experiments. Therefore,
only the West and East coast conventions are
allowed.
It is interesting to note that, for both
conventions, the dimensions
$n$ and $m$ are the same.
Indeed, in the West coast convention
$(p,q)=(1,3)$ and $n=p-q=6\mod 8$, while
for the East coast convention
$(p,q)=(3,1)$ and $n=q-p=6\mod 8$.
In other words, the $KO$ dimension 
$n$ is the number of time dimensions $t$ minus
the number of space dimensions $s$ while
$m$ is their sum $t+s$ whatever the convention. 
It is therefore tempting to intepret 
the dimension $m$ and $n$ in terms of 
time and space dimensions by solving
$n=t-s\mod 8$ and $m=t+s\mod 8$.
This associates a space dimension and a time
dimension to any Clifford algebra.

\begin{table}
\centering
\begin{tabular}{|l||l|l|l|l|}
\hline
$m$, $n$ & 0 & 2 & 4 & 6\\
\hline
\hline $\kappa, \epsilon$  & 1 & -1 & -1 & 1\\
\hline $\kappa'', \epsilon''$  & 1 & -1 & 1 & -1\\
\hline
\end{tabular}
\caption{Signs $\epsilon,\kappa$ and $\epsilon'',\kappa''$ in terms of
the KO-dimension $n$ and $m$.
\label{tabeps}
  }
\end{table}

%\begin{table}
%\centering
%\begin{tabular}{|l||c|c|c|c|}
%\hline
%Convention & $p_0$ & $q_0$ & $J$ & $\eta$ \\
%\hline
%\hline West-coast  & $\frac{m+n}{2}$  & $\frac{m-n}{2}$ & $J_-$ & $\eta_+$ \\
%\hline East-coast  & $\frac{m-n}{2}$  & $\frac{m+n}{2}$ & $J_+$ & $\eta_-$ \\
%\hline South-coast  & $\frac{n-m}{2}$  & $\frac{-m-n}{2}$ & $J_-$ & $\eta_-$ \\
%\hline North-coast  & $\frac{-m-n}{2}$  & $\frac{n-m}{2}$ & $J_+$ & $\eta_+$ \\
%\hline
%\end{tabular}
%\caption{Value of $p$, $q$, $J$ and $\eta$ for each convention.
%The allowed values of $p$ and $q$ corresponding to a
%given pair $(p_0,q_0)$ are $p=p_0+4i+8j$ and 
%$q=q_0+4i+8k$, where $p$ and $q$ are nonnegative integers,
%$i$, $j$, $k$ are integers.
%\label{tabconv}
%  }
%\end{table}

\begin{table}
\centering
\begin{tabular}{|l||c|c|c|c|}
\hline
Convention & $m$ & $n$ & $J$ & $\eta$ \\
\hline
\hline West-coast  & $p+q$  & $p-q$ & $J_-$ & $\eta_+$ \\
\hline East-coast  & $p+q$  & $q-p$ & $J_+$ & $\eta_-$ \\
\hline North-coast  & $-p-q$  & $p-q$ & $J_-$ & $\eta_-$ \\
\hline South-coast  & $-p-q$  & $q-p$ & $J_+$ & $\eta_+$ \\
\hline
\end{tabular}
\caption{Value of the KO-dimension $n$ and of
$m$ and operators as a function of convention.
The dimensions $m$ and $n$ are to be
understood modulo 8.
\label{tabconv}
  }
\end{table}

%The three operators just defined are related with natural automorphisms of the complexified Clifford algebra. By its very definition $J$ implements by its adjoint action $u\mapsto JuJ^{-1}$ the antilinear automorphism $c$. The chirality 
%$\chi$ implements the main automorphism of 
%the Clifford algebra:
%$\chi u \chi=u$ (resp. $\chi u \chi=-u$)
%if $u$ is the product of an even (resp. odd) number
%of gamma matrices. Finally there is  the reversion (or transpose) anti-automorphism
%(i.e. $\gamma_{i_1}\dots\gamma_{i_k}\mapsto
%\gamma_{i_k}\dots\gamma_{i_1}$), and it turns out that the composition of the main automorphism, real structure and reversion is implemented 
% by
%$u\mapsto \eta u^\dagger \eta$.

\begin{table}
\centering
\begin{tabular}{|l|l|l|l|l|}
\hline
\backslashbox {$m$}{$n$} & 0 & 2 & 4 & 6\\
\hline 0 & (0,0) (4,4) & (1,7) (5,3) & (2,6)  (6,2)& (3,5) (7,1)\\
\hline 2 & (1,1) (5,5) & (2,0) (6,4) & (3,7) (7,3) & (0,2) (4,6)\\
\hline 4 & (2,2) (6,6) & (3,1) (7,5) & (4,0) (0,4) & (1,3) (5,7)\\
\hline 6 & (3,3) (7,7) & (4,2) (0,6) & (5,1) (1,5) & (6,0) (2,4)\\
\hline
\end{tabular}
\caption{$(t,s)$ where $t$ is the number of time dimensions and
$s$ the number of space dimensions $s$ as a function of 
$m$ and $n$.  The general solution is
$(t+8j,s+8k)$, where $j$ and $k$ are integers.
The relation between $(p,q)$ and $(t,s)$ is:
$p=t$, $q=s$ (West coast),
$p=s$, $q=t$ (East coast),
$p=-s$, $q=-t$ (North coast) and
$p=-t$, $q=-s$ (South coast).
\label{tabst}
  }
\end{table}

By inverting the relation between
$(s,t)$ and $(m,n)$, we 
can associate \emph{two} pairs of space and time dimensions
$(j,k)$ modulo 8 to every pair $(m,n)$
(see Table~\ref{tabst}).
Indeed, if $(j,k)$ is a solution of
$j-k=n\mod 8$ and $j+k=m\mod 8$,
then $(j+4,k+4)$ is also a solution.
This corresponds to the Clifford algebra isomorphism
$C\ell(s,t+8)\simeq C\ell(s+8,t)\simeq C\ell(s+4,t+4)$~\cite{Garling}. %p.111
The relation between $(s,t)$ and $(p,q)$ for the
four conventions is given in the caption of
Table~\ref{tabst}.

\section{Generalization}
\label{general-sect}
We generalize the previous results by 
defining a \emph{mod-8-spacetime 
representation} to be a quadruple
$S=(\calH,\chi,\eta,J)$, where $\calH$ is
a complex Hilbert space equipped with
two self-adjoint involutions
$\chi$ and $\eta$ (i.e.  $\chi^2=\eta^2=1$)
and an anti-unitary operator $J$
that satisfy Eqs.~\eqref{Jdeux} 
to~\eqref{etachi}  for some
signs $\epsilon$, $\epsilon''$, $\kappa$
and $\kappa''$.
%We denote $\sigma(S)=(\epsilon,\epsilon'',\kappa,\kappa')$.
%The map $\phi$ associates
%space and time dimensions
%to each mod-8-spacetime representation.
%However, as for the Brauer-Wall group~\cite{Lam-05},
%this assignment can only be meaningful if
%it is compatible with the graded tensor product
%that we define now.
The $KO$-dimension $n$ is compatible with the tensor
product, in the sense that the dimension of 
the tensor product of two algebras is the sum
of their dimensions.
It is physically important that the same holds for the
other dimension $m$, because the state space of
a many-body problem is built from the tensor product
of one-particle states.

By using the chirality operator $\chi$,
we can write $\calH=\calH_+\oplus\calH_-$, where
$\chi v = \pm v$ for $v\in\calH_\pm$.
An element $v$ of $\calH_\pm$ is said to be homogeneous
and its parity is $|v|=0$ if $v\in \calH_+$
and $|v|=1$ if $v\in \calH_-$.
The parity of a linear or antilinear map $T$ on $\calH$
is $|T|=0$ if $\chi T \chi = T$ and
$|T|=1$ if $\chi T \chi = -T$.
From relations~\eqref{Jchi} and \eqref{etachi} we see that
$\epsilon''=(-1)^{|J|}$ and
$\kappa''=(-1)^{|\eta|+|J|}$.

The graded tensor product $\hat\otimes$
of operators is defined by
$(T_1\hat\otimes T_2)(\phi_1\otimes\phi_2)=
(-1)^{|T_2||\phi_1|}  T_1 \phi_1 \otimes T_2 \phi_2$
when $\phi_1$ and $\phi_2$ are homogeneous.
It is the natural tensor product of
Clifford algebra theory thanks to
Chevalley's relation~\cite{Lawson}:
\begin{eqnarray}
C\ell(p_1,q_1)\hat\otimes C\ell(p_2,q_2)
 &=& C\ell(p_1+p_2,q_1+q_2),
\label{Chevalley}
\end{eqnarray}
which shows that the graded tensor product
is indeed compatible with space and
time dimensions.

The graded 
tensor product $T_1\hotimes T_2$
is the same operator as the
non-graded tensor product
$T_1 \chi_1^{|T_2|}\otimes T_2$. 
For example, 
the graded Dirac operator of the
tensor product:
$D=D_1\hotimes 1 + 1 \hotimes D_2$ 
is the same operator as the
one given by Connes:
$D=D_1\otimes 1 + \chi_1 \otimes D_2$.

Let us consider two mod-8-spacetime representations
$S_1=(\calH_1,\chi_1,\eta_1,J_1)$ and
$S_2=(\calH_2,\chi_2,\eta_2,J_2)$
with signs determined by $(m_1,n_1)$ and
$(m_2,n_2)$, respectively.
Then, the graded tensor product
$S=S_1\hat\otimes S_2$ is the mod-8-spacetime
representation defined by the Hilbert space
$\calH=\calH_1\otimes\calH_2$ and the operators
\begin{eqnarray}
\chi &=& \chi_1 \hat\otimes \chi_2,\label{tensorchi}\\
J &=&  J_1 \chi_1^{|J_2|} \hat\otimes J_2 \chi_2^{|J_1|},
  \label{tensorJ}\\
\eta &=&  i^{|\eta_1||\eta_2|}
   \eta_1 \chi_1^{|\eta_2|}
   \hat\otimes
   \eta_2 \chi_2^{|\eta_1|}.\label{tensoreta}
\end{eqnarray}
To derive these equations, we first define
an indefinite inner product
$(\cdot,\cdot)$ on $\calH$ in terms of the 
indefinite inner products
$(\cdot,\cdot)_1$ and $(\cdot,\cdot)_2$ 
on $\calH_1$ and $\calH_2$. 
The adjoint of an operator $T$ with respect to an indefinite
inner product is denoted by $T^\times$.

We consider the example where
$\calH_1$ and $\calH_2$ are the spinor spaces of
two Clifford algebras. Then, we use
Robinson's theorem~\cite{Robinson-88}, which states that
there is a unique indefinite inner product
(up to a real scalar factor)
on the space of spinors 
such that the $\gamma^\mu$ matrices generating the
Clifford algebra satisfy 
$(\gamma^\mu)^\times=\gamma^\mu$.
If we impose that $\eta_+$ was chosen 
for both the first and second Clifford algebras,
we obtain the following indefinite inner product
on $\calH$:
\begin{eqnarray*}
(\phi_1\otimes \phi_2,\psi_1\otimes\psi_2) &=& i^{|\eta_1||\eta_2|}
   (\phi_1,\psi_1)_1 (\phi_2,\chi_2^{|\eta_1|}
   \psi_2)_2.
\end{eqnarray*}
Since this definition depends only on
$|\eta_1|$, $|\eta_2|$ and $\chi_2$, we can
extend it to any mod-8-spacetime representation.
Formula~\eqref{tensoreta} expresses
the compatibility of $\eta$ with this
indefinite inner product.
This ensures
\begin{eqnarray*}
\langle\phi_1\otimes \phi_2,\psi_1\otimes\psi_2\rangle
&=&  \langle\phi_1,\psi_1\rangle_1 
\langle\phi_2, \psi_2\rangle_2,
\end{eqnarray*}
and implies the Kasparov identities~\cite{Kasparov-81}
\begin{eqnarray*}
(T_1\hat\otimes T_2)^\times &=& (-1)^{|T_1||T_2|}
   T_1^\times \hat\otimes T_2^\times,\\
(T_1\hat\otimes T_2)^\dagger &=& (-1)^{|T_1||T_2|}
   T_1^\dagger \hat\otimes T_2^\dagger,
\end{eqnarray*}
for the tensor product of two linear operators
and
\begin{eqnarray*}
(T_1\hat\otimes T_2)^\times &=& (-1)^{|\eta_1||\eta_2|+|T_1||T_2|}
   T_1^\times \hat\otimes T_2^\times,\\
(T_1\hat\otimes T_2)^\dagger &=& (-1)^{|T_1||T_2|}
   T_1^\dagger \hat\otimes T_2^\dagger,
\end{eqnarray*}
for the tensor product of two antilinear operators.

The tensor product of mod-8-spacetime representations
is associative and symmetric~\cite{Bizi-PhD}.
It can be checked that
$\chi$ and $\eta$ are self-adjoint involutions
and $J$ is an anti-unitary map, which satisfy
Eqs~\eqref{Jdeux} to \eqref{etachi} for the signs of
some dimensions $(m,n)$.
Indeed,
we first observe through an explicit calculation
that the signs associated to 
$S_1\hat\otimes S_2$ only depend
on the signs associated to $S_1$ and $S_2$:
$\epsilon =(-1)^{|J_1||J_2|} \epsilon_1\epsilon_2$
(where $(-1)^{|J_1||J_2|}  =
(1+\epsilon''_1+\epsilon''_2-\epsilon''_1\epsilon''_2)/2)$,
$\epsilon''=\epsilon''_1\epsilon''_2$,
$\kappa=(-1)^{(|\eta_1|+|J_1|)(|\eta_2|+|J_2|)}\kappa_1\kappa_2$
and
$\kappa''=\kappa''_1\kappa''_2$.
Then, the additivity of the dimensions in 
mod-8-spacetime representations follows from
the fact that it holds for Clifford algebras
(see Eq.~\eqref{Chevalley}).
%Now let $A_1$ and $A_2$ be two Clifford algebras such that
%$\sigma(A_i)=\sigma(S_i)$, $i=1,2$. Then : $\phi(\sigma(S_1\otimes
%S_2))=\phi(\sigma(A_1\otimes A_2))$ since  $\sigma(S_1\otimes S_2)$
%only depends on $\sigma(S_1)=\sigma(A_1)$ and $\sigma(S_2)=\sigma(A_2)$.
%However the space and time dimensions are additive for Clifford
%algebras, hence $\phi(\sigma(A_1\otimes
%A_2))=\phi(\sigma(A_1))+\phi(\sigma(A_2))=\phi(\sigma(S_1))+\phi(\sigma(S_2))$.
% Thus this additive property extends to all 
%mod-8-spacetime representations as
%$m=m_1+m_2\mod 8$ and $n=n_1+n_2\mod 8$.

%Extending this classification to odd-dimensional
%spaces appears non-trivial, let alone because
%the main automorphism of an odd-dimensional
%Clifford algebra is not inner~\cite{Budinich-16}
%(in other words, there is no $\chi$).

These space and time dimensions can be used
to classify topological insulators and superconductors
with symmetries, as well as for the investigation
of $PT$-symmetric Hamiltonians, because
of the presence of a 
Krein-space structure~\cite{Mostafazadeh-02-1,Tanaka-06,Mannheim-16},
to which we now turn 
because it is crucial to generalize spectral triples
to pseudo-Riemannian manifolds.

%\pink{Note that, according to our classification, equivalent
%Clifford algebras do not belong to the same class.
%For example, $C\ell(3,0)$ and $C\ell(1,2)$ are
%algebraically equivalent but correspond to
%$(m,n)=(3,3)$ and $(3,7)$, respectively. 
%This is due to the fact that 
%$(\chi,J,\eta)$ are not preserved by the algebraic
%isomorphism. Similar refined isomorphisms of Clifford
%algebras (e.g. geometric equivalence) have been proposed 
%in the literature~\cite{Botman-06}.
%}

\section{Krein spaces}
We already met Hilbert spaces $\calH$,
with scalar product denoted $\langle\cdot,\cdot\rangle$,
which are equipped with a fundamental
symmetry (i.e. a self-adjoint operator 
$\eta$ such that $\eta^2=1$) defining an indefinite
inner product $(\phi,\psi)=\langle \phi,\eta\psi\rangle$.
In the mathematical literature, such a space is
called a \emph{Krein space}.
It was noticed by Helga Baum~\cite{Baum-81} that 
the spinor bundle of a pseudo-Riemannian 
manifold is naturally equipped with the structure
of a Krein space.
This was generalized by Alexander
Strohmaier to noncommutative geometry~\cite{Strohmaier-06}.

Substituting an indefinite inner product for a scalar product
has a striking physical consequence:
the possible existence of states
with negative ``probabilities'' (i.e. such that $(\psi,\psi)<0$).
These states were first met in physics by Dirac in 1942 in his quantization of
electrodynamics~\cite{Dirac-42}.
He interpreted negative-norm states as 
describing a \emph{hypothetical world}~\cite{Kragh}.
Negative-norm states have now become familiar
in physics through their role in the Gupta-Bleuler
and Becchi-Rouet-Stora-Tyutin (BRST)
quantizations of gauge fields.

In most applications, the indefinite inner product 
$(\cdot,\cdot)$ is natural (i.e. uniquely defined,
up to a scalar factor, by some symmetry condition)
and the scalar product $\langle \cdot,\cdot\rangle$ 
is somewhat arbitrary. In a Lorentzian manifold,
the scalar product corresponds to the Wick
rotation following 
some choice of a time-like direction
(see Ref.~\cite{Dungen-13} for a precise definition).
Krein spaces are a natural framework
for gauge field theories~\cite{Horuzhy-89,Nakanishi,Strocchi-13}
and Lorentzian spectral 
triples~\cite{Dungen-15,Strohmaier-06,Dungen-13,Suijlekom-04,Paschke-06,%
Marcolli-08,Besnard-17}.

We present now some essential properties of 
operators on Krein spaces but, true
to the physics tradition, we do not
describe their functional analytic properties.
If $\calK$ is a Krein space, a
linear operator $T:\calK\to\calK$ 
has a Krein-adjoint $T^\times$ defined by
$(T^\times x,y)=(x,Ty)$ for every $x$
and $y$ in $\calK$.
A linear operator is Krein-self-adjoint
if $T^\times=T$ and Krein-unitary 
if $T^\times T=T T^\times = 1$. 

An anti-linear map
(i.e. a map $T:\calK\to\calK$ 
such that $T(\alpha x + \beta y)
=\overline{\alpha} Tx + \overline{\beta} Ty$)
has a Krein-adjoint $T^\times$ defined
by $(y,T^\times x)=(x,T y)$.
It is Krein-anti-unitary if, furthermore,
 $T^\times T=T T^\times =1$.

Any fundamental symmetry $\eta$ is Krein self-adjoint.
The relation between the Krein adjoint
$T^\times$ and the adjoint $T^\dagger$
with respect to the scalar product of $\calH$ is
$T^\dagger = \eta T^\times \eta$.

In physical applications the Krein adjoint
is the most natural. For example, the Dirac
operator on a pseudo-Riemannian manifold is
Krein-self-adjoint. In Gupta-Bleuler
quantization the Krein adjoint
is covariant. In gauge field theory, the 
BRST charge is Krein self-adjoint.
The Hilbert adjoint depends on the choice of
a fundamental symmetry, it is not covariant
in Gupta-Bleuler quantization and the BRST approach.
We can now define an indefinite spectral triple.

\section{Indefinite spectral triple}
\label{ind-sect}
After some pioneering 
works~\cite{Lizzi-97,Gracia-98,Elsner-99},
many papers were devoted to the extension of 
noncommutative geometry to Lorentzian 
geometry~\cite{Dungen-15,Strohmaier-06,Dungen-13,%
Suijlekom-04,Paschke-06,Marcolli-08,%
Barrett-07,Kopf-02,Moretti-03-2,Paschke-04,%
Bieliavsky-04,Kopf-06,%
Jureit-07,Paschke-07,%
Bieliavsky-08,Besnard-09,Borris-10,Nelson-10,Verch-11,%
Franco-13,Franco-14-2,Franco-14-3,%
Besnard-15,Franco-15,Andrea-Lizzi-16,Dungen-16,Eckstein-16,%
Franco-16,Watcharangkool-17,Devastato-17}.
Inspired by these references, we define
an even-dimensional real indefinite spectral triple to be:
\begin{enumerate}
\item A $*$-algebra $\calA$ represented on a
 Krein space $\calK$ equipped with a
Hermitian form $(\cdot,\cdot)$
and a fundamental symmetry $\eta$.
We assume that the representation satisfies
$\pi(a^*)=\pi(a)^\times$.
\item A chirality operator $\chi$, i.e. a linear map 
on $\calK$ such that 
$\chi^2=1$ and  $\chi^\dagger=\chi$
(where the adjoint $\chi^\dagger$ is defined
by $\chi^\dagger=\eta\chi^\times \eta$).
The algebra commutes with $\chi$.
\item An antilinear  charge conjugation $J$, 
  such that $J^\dagger J=1$.
\item A set of signs $(\epsilon,\epsilon'',\kappa,\kappa'')$
  describing relations \eqref{Jdeux} to \eqref{etachi}
   between $\chi$, $\eta$ and $J$.
\item A Krein-self-adjoint Dirac operator $D$, which 
satisfies $JD=DJ$ and $\chi D = - D\chi$.
\end{enumerate}
The hypothesis $\pi(a^*)=\pi(a)^\times$ is a simplifying assumption, which
is well-adapted to particle physics applications, but which needs not be
longer true when discrete structures replace 
manifolds~\cite{Besnard-17-II}.
We refer the reader to 
Refs.~\cite{Strohmaier-06,Dungen-13,Franco-14-2,Dungen-16} for the
functional analytic aspects of indefinite
spectral triples.
If we compare with Connes' spectral triples,
we see that we have an additional object
(the fundamental symmetry $\eta$) and two
additional signs: $\kappa$ and $\kappa''$.
Because of this more complex structure,
the $KO$-dimension $n$ is no longer enough to classify
indefinite spectral triples and we need
both $m$ and $n$.
The classification carried out in section~\ref{general-sect}
holds also for indefinite spectral triples
because they are particular cases of
mod-8-spacetime representations.
More precisely, Let
$(\calA_1,\calH_1,D_1,J_1,\chi_1,\eta_1)$
and
$(\calA_2,\calH_2,D_2,J_2,\chi_2,\eta_2)$ 
be two real even-dimensional indefinite spectral
triples. Supplement the tensor products defined 
in section~\ref{general-sect}  with 
\begin{eqnarray*}
\calA &=& \calA_1 \hat\otimes \calA_2,\\
D &=& D_1 \hat\otimes 1 + 1 \hat\otimes D_2,\\
\pi &=& \pi_1 \hat\otimes \pi_2.
\end{eqnarray*}

It can be checked that this tensor product 
is indeed a real even-dimensional indefinite
spectral triple
(i.e. $D^\times = D$ and $D$ commutes with $J$
and anticommutes with $\chi$).

The extension of this tensor product to odd-dimensional 
indefinite spectral triples seems difficult, if one 
considers the complexity of the
Riemannian case~\cite{Vanhecke-99,Dabrowski-11,Vanhecke-12,%
Cacic-13,Farnsworth-16}. Note that Farnsworth
also advocates the use of a graded tensor product~\cite{Farnsworth-16}.

We define now the indefinite spectral triple 
encoding models of particles physics.

\section{Particle physics models}
Particle physics models (QED, electroweak, standard model)
are described by an almost commutative
spectral triple, i.e. the tensor product of
the spectral triple $\calS_1$  of a manifold $M$ and
a finite dimensional spectral triple $\calS_2$.

Connes and Lott~\cite{Connes-90} derived the
fermionic and gauge Lagrangians of the standard model
in Riemannian space. In Lorentzian spacetime
Dungen found the fermionic Lagrangian but
the gauge Lagrangian
is considered to be an open problem~\cite{Dungen-15}.
Surprisingly, the problem was already solved
in Elsner's outstanding Master's thesis~\cite{Elsner-99}
(see also \cite{Elsner-01,Elsner-02-weak,Elsner-02-SM}),
where many aspects of noncommutative geometry
(e.~g. Connes differential algebra, curvature, bosonic 
and fermionic Lagrangians)
are generalized to the case where the Hilbert space is
replaced by a vector space equipped with a 
sesquilinear form. Indefinite spectral triples clearly
fit into that framework.
When the base manifold $M$ is not compact,
Elsner defines a family of bosonic Lagrangians
$A_n = \int_{U_n} \sqrt{|g|}dx \calL_b(x)$,
%\begin{eqnarray*}
%A_n &=& \int_{U_n} d_\mu \calL,
%\end{eqnarray*}
where the relatively compact open sets $U_n$
form an exhausting family covering $M$
and $\calL_b = -\Tr(\theta^\times \theta)$
is the Lagrangian density.
The two-form $\theta$ in Connes' differential
algebra is the curvature of the gauge potential $\rho$
and the trace is
\begin{eqnarray*}
\Tr(M_1\otimes M_2) &=& \Tr_1(M_1)\Tr_2(M_2),
\end{eqnarray*}
where $\Tr_1$ is the trace over the spinor fibre over $M$
and $\Tr_2$ the trace over the finite dimensional Krein space
$\calK_2$. 

The family of open sets ensures that,
for any compactly supported variation,
there is an $n_0$ such that $U_n$
contains the support of the variation for every
$n>n_0$ and the variation of $A_n$ 
is well defined. 
This is compatible with the NCG point of view because
Suijlekom proved the conceptually important fact that
any noncommutative geometry can be considered
as an algebra bundle over a Hausdorff 
base space~\cite{Suijlekom-16}. We calculate
$\calL_b$ for QED in this paper and for
the electroweak and standard models in a
forthcoming publication.
We use the
fermionic Lagrangian
$\calL_f=(\Psi,D\Psi)$~\cite{Dungen-15,Elsner-99,Barrett-07}.
Note that in the Riemannian NCG Lagrangian 
the first $\Psi$
is replaced by $J\Psi$~\cite{Connes-Marcolli}.

An important and long-standing 
problem is that 
only \emph{physical} states $\Psi$
must be used in the fermionic Lagrangian,
while the trace $\Tr$ is over the whole
space $\calK_1\otimes\calK_2$.
Lizzi and coll.~\cite{Lizzi-97} pointed out that the bosonic
Lagrangian over physical states
is different from $\calL_b$ and not physically valid
(it contains $CPT$-symmetry violating terms).
We discuss now this so-called 
\emph{fermion doubling} problem.

\section{Fermion doubling problem}
The name of the fermion doubling (or quadrupling)
problem comes from the fact that 
a state in $\calK_1\otimes\calK_2$
corresponding to a particle $p$ can
be written as a linear combination of $\psi\otimes p_L$,
$\psi\otimes p_R$, $\psi\otimes p^c_L$
and $\psi\otimes p^c_R$.
Since $\psi\in \calK_1=\Gamma(M,S)$ is a four-dimensional
(Dirac) spinor, each particle is described
by a 16-dimensional vector instead of a
four-dimensional one.

The solution proposed by 
Elsner and coll.~\cite{Elsner-99,Elsner-02-SM} 
(which is different from the one of Ref.~\cite{Lizzi-97}) is 
to consider as physical the linear combinations
of $\psi_L\otimes p_L$, $\psi_R\otimes p_R$,
$\psi^c_L\otimes p^c_L$ and 
$\psi^c_R\otimes p^c_R$, where
$\psi_L$ and $\psi_R$ are two-dimensional
Weyl fermions (obtained as
$\chi^\pm_1\psi$ where
$\chi^\pm_1=(1\pm\chi_1)/2$),
$\psi^c_{L/R}=J_1\psi_{L/R}$ in the fermionic 
space $\calK_1$ and
$p^c_{L/R}=J_2 p_{L/R}$ in the gauge space $\calK_2$.
The degrees of freedom are reduced to four
(two for $\psi_L\otimes p_L$ and two for 
$\psi_R\otimes p_R$). The antiparticles states do not
correspond to additional degrees of freedom because
they are obtained by applying $J$ to the particle 
states~\cite{Elsner-99,Lazzarini-01}.

We propose two justifications for this choice.
Lizzi and coll.~\cite{Lizzi-97}
noticed that left-handed particle (for the Lorentz group) $\psi_L$
must also be left-handed particles $p_L$ for the gauge group
to be physically meaningful.
We can complete this by pointing that
on-shell particles and anti-particles 
are solutions of different (i.e. charge conjugate)
Dirac equations in the presence of an external field.
Therefore, particles and antiparticles are distinct
and there is no reason to couple $\psi^c_{L/R}$ with $p_{R/L}$.

The second justification comes from Grand Unified Theories (GUT),
where each particle $p$ also appears in 
four varieties: $p_L/p_R$ and 
$p^c_L/p^c_R$~\cite{Baez-10}. 
%Note that $p^c_L=J p_L$ is right-handed
%because $J$ inverts parity.
In GUT, the identification of the spinor and gauge variables
is so obvious that it is implicit and a state
like $\psi_L \otimes p_L$ is simply denoted by $p_L$.

All the physical states satisfy 
the Weyl property $\chi\Psi=\Psi$ proposed long ago by
Connes~\cite{Connes-Bourbaki}.
It was also proposed to
supplement the Weyl condition with the
Majorana condition $J\Psi=\Psi$~\cite{Barrett-07}.
However, this would allow states of the form
$\psi^c_R\otimes p_L+\psi_R\otimes p^c_L$,
which are not physical from our point of view,
and the selected space of states would not be a complex
vector space because $J$ is antilinear.

We can now define the trace of 
$M=M_1\otimes M_2$ over physical states
as
\begin{eqnarray*}
\Tr' M &=& \sum_{P=L/R}
   \sum_{\psi_P,p_P}
 \big(  \langle \psi_P,M_1\psi_P\rangle 
      \langle p_P,M_2 p_P\rangle
\\&&
 +  \langle \psi^c_P,M_1\psi^c_P\rangle 
      \langle p^c_P,M_2 p^c_P\rangle\big).
\end{eqnarray*}
Note that $\Tr$ and $\Tr'$ are different
because $\Tr$ has no sum over
$\psi^c_P$ and has sums over non physical
states such as
$\psi_L\otimes p_R$ or $\psi_L\otimes p^c_R$.
To rewrite this in a more convenient way,
we can use the operator $\varpi$ defined
by Connes~\cite{Connes-95-reality}
(who calls it $\epsilon$) and identified
by Elsner as a particle/antiparticle operator.
In our framework, $\varpi$ commutes with 
$\calA_2$, $\chi_2$ and $\eta_2$
and anticommutes with $J_2$. Connes also requires
$\varpi$ to commute with $D_2$, but we shall not 
use this property because it would cancel the 
Majorana mass of the neutrinos.
We define $\varpi$ by $\varpi p_{L/R}=p_{L/R}$ and
$\varpi p^c_{L/R}=-p^c_{L/R}$.
By using the duality property of anti-linear operators
$\langle J_i u,v\rangle= \langle J_i^\dagger v,u\rangle$ for
$u$ and $v$ in $\calK_i$,
we can transform $\Tr'$ into
\begin{eqnarray*}
\Tr' M &=& \sum_{\sigma=\pm} \big(
    \Tr_1(M_1\chi_1^\sigma) \Tr_2(M_2\chi_2^\sigma\varpi^+)
\\&& +
    \Tr_1(J_1M_1^\dagger J_1^{-1} \chi_1^{\sigma}) 
   \Tr_2(J_2 M_2^\dagger J_2^{-1} \chi_2^{\sigma}\varpi^+)\big),
\end{eqnarray*}
where $\varpi^+=(1+\varpi)/2$ projects $\calK_2$
onto particle states.

This choice of physical states solves
the fermion doubling problem because,
although $\Tr$ and $\Tr'$ are generally not
proportional,
 $\Tr \theta^\times \theta=2 \Tr'\theta^\times \theta$ for 
the indefinite spectral triples of QED, the electroweak
and standard models. This will be clear for
QED and will be discussed for the other models in
a forthcoming paper.

\section{Indefinite spectral triple of QED}
The NCG model of QED in Riemannian space
time was described by Dungen and
Suijlekom~\cite{Dungen-13-QED} and in
Lorentzian space by Dungen~\cite{Dungen-15}
who did not include antiparticles and
the charge conjugation operator.
The manifold spectral triple consists of
a Lorentzian 4D manifold $M$, a Dirac operator
$D_1=i\gamma^\mu \nabla_\mu$, where 
$\nabla_\mu$ is the spin connection 
and the Krein space $\Gamma(M,S)$~\cite{Baum-81}.
In the chiral representation of the
gamma matrices~\cite{Itzykson} the operators are
\begin{eqnarray*}
\chi_1 = {\sixj{-1}{0}{0}{1}},
\eta_1 = \sixj{0}{1}{1}{0},
J_1 = i \sixj{0}{\sigma^2}{-\sigma^2}{0} K,
\end{eqnarray*}
where $K$ means complex conjugation
and $\sigma^2$ is the Pauli matrix.
We use the same algebra $\calA=\bbC\oplus \bbC$
as Dungen~\cite{Dungen-15} and 
the Krein space $\calK_2=\bbC^4$ 
with basis states $(e_L,e_R,e_R^c,e_L^c)$.
The representation and operators are,
in terms of Pauli matrix $\sigma^1$,
$\sigma^2$ and $\sigma^3$
\begin{eqnarray*}
\pi_2(a,b)  &=& \left(\begin{array}{cc}
  a1 & 0\\
   0 & b1
   \end{array}\right),\quad
\chi_2  = - \left(\begin{array}{cc}
  \sigma^3 & 0\\
   0 & \sigma^3
   \end{array}\right),\\
\eta_2  &=& \left(\begin{array}{cc}
  -\sigma^3 & 0\\
   0 & \sigma^3
   \end{array}\right),\quad
J_2  = \left(\begin{array}{cc}
 0 & \epsilon_2 \sigma^1\\
   \sigma^1  & 0
   \end{array}\right)K,\\
\varpi  &=& \left(\begin{array}{cc}
  1 & 0\\
   0 & -1
   \end{array}\right),\quad
D_2  = im \left(\begin{array}{cc}
   -\sigma^2 & 0\\
   0 & \sigma^2
   \end{array}\right).
\end{eqnarray*}
To compare the fermionic Lagrangian on 
particles and antiparticles we use the relation
$(J\Psi,DJ\Psi)=\epsilon\kappa (\Psi,D\Psi)$.
We determine $\epsilon\kappa$ by computing the
dimensions of the total spectral triple.
For the manifold spectral triple
$(m_1,n_1)=(4,6)$ and for the
finite spectral triple 
$(m_2,n_2)=(2,2)$ if $\epsilon_2=-1$ and
$(m_2,n_2)=(6,6)$ if $\epsilon_2=1$. 
In the literature it is generally assumed
that $n_2=6$ but $n_2=2$ is also valid for QED. 
The total spectral triple has now dimensions
$(m,n)=(6,0)$ if $\epsilon_2=-1$ and
$(m,n)=(2,4)$ if $\epsilon_2=1$.
In both cases $\epsilon\kappa=1$, which implies
$(J\Psi,DJ\Psi)=(\Psi,D\Psi)$.
Thus, we can also define the fermionic Lagrangian over the
particle states only.

The physical Lagrangian for QED with a massive electron is:
\begin{eqnarray*}
\calL &=& -\frac14 F^{\mu\nu}F_{\mu\nu}
  + (\psi,\big(i\gamma^\mu 
    (\nabla_\mu + i q A_\mu) -m\big)\psi),
\end{eqnarray*}
where $q<0$ is the electron charge
and $F_{\mu\nu}=\partial_\mu A_\nu-\partial_\nu A_\mu$.

The Dirac corresponding to this triple is
\begin{eqnarray*}
D(A) &=& D - q \gamma^\mu A_\mu\otimes \varpi,
\end{eqnarray*}
where 
\begin{eqnarray*}
-qA_\mu &=& i \sum_j \pi(a_j) \partial_\mu \pi(b_j),
\end{eqnarray*}
is self-adjoint.
The curvature is 
\begin{eqnarray*}
\theta &=& -\frac{iq}{2} \sum_{\mu\nu}
(\gamma^\mu\gamma^\nu-g^{\mu\nu})F_{\mu\nu} \otimes \varpi,
\end{eqnarray*}
and the bosonic Lagrangian is
\begin{eqnarray*}
\calL_b &=& -8q^2 F_{\mu\nu} F^{\mu\nu}.
\end{eqnarray*}
The prefactor of $\calL_b$ is not correct.
This problem was elegantly 
solved~\cite{Iochum-95,Elsner-99,Elsner-01}
by choosing a positive definite 
$z$ which commutes with $\pi(\calA)$,
$J \pi(\calA) J^{-1}$ and $D$ 
to redefine the trace
as $\Tr_z(M)=\Tr(zM)$. In the case of 
QED a solution of the constraints is
$z=\rho 1\otimes 1$, where $\rho>0$.
We can now define the Lagrangian
\begin{eqnarray*}
\calL_{\mathrm{CLE}} &=& \Tr_z (\theta^\times \theta) 
 + \frac12 (\Psi,D(A)\Psi),
\end{eqnarray*}
which we call the Connes-Lott-Elsner Lagrangian since it was
originally proposed by Connes and Lott in the Riemannian case, and then
extended by Elsner to general signatures. 
Observe that, since noncommutative 2-forms are defined modulo the 
junk~\cite{Connes94},
the expression $\Tr(\theta^\times \theta)$ does not have an
immediate meaning. 
In the Riemannian case, this problem is solved by
projecting $\theta$ onto the orthogonal of the junk.
This solution can be applied in general signature 
if and only if the restriction of the 
indefinite inner product to the junk is non degenerate.
Remarkably, this property holds in QED as
well as in the full standard 
model~\cite{Elsner-99,Bizi-PhD}.

The QED bosonic Lagrangian becomes
$\calL_b = -8\rho q^2 F_{\mu\nu} F^{\mu\nu}$
and we can choose $\rho=1/(32q^2)$ to obtain
the physical bosonic Lagrangian.
The fermionic Lagrangian $\frac12 (\Psi,D(A)\Psi)$
with 
\begin{eqnarray*}
\Psi &=& 
  (1+J)(\chi_1^+\psi\otimes e_R + \chi_1^-\psi\otimes e_L),
\end{eqnarray*}
where $\langle\psi,\psi\rangle=1$, becomes
\begin{eqnarray*}
\calL_f &=& 
 (\psi,\big(i\gamma^\mu 
    (\nabla_\mu + i q A_\mu) -m\big)\psi),
\end{eqnarray*}
and the CLE-Lagrangian exactly coincides with the physical one.

To complete this section, we remark that 
particles and antiparticles have the same
mass and opposite currents.
In textbook QED, these properties are
achieved through the anticommutativity
of the normal product of fermion operators~\cite{Itzykson}.
In our framework, this property follows from the
fundamental symmetry.
Indeed, by using 
$J_1\gamma^\mu=-\gamma^\mu J_1$ and 
$(\gamma^\mu)^\times=\gamma^\mu$
we get
\begin{eqnarray*}
\langle J\Psi,\eta J\Psi\rangle &=& \langle\Psi,\eta \Psi\rangle,\\
\langle J\Psi,\eta (\gamma^\mu\otimes 1)J\Psi\rangle &=& 
-\langle\Psi,\eta(\gamma^\mu\otimes 1) \Psi\rangle,
\end{eqnarray*}
which means that the mass is conserved and the current is
reversed by charge conjugation. We did not use 
any anticommutation relation.

\section{Conclusion}

A particularly appealing aspect of noncommutative
geometry is that
the internal (fibre) and external (manifold) degrees
of freedom are put into a common geometric framework.
Real Clifford algebras can also unify spacetimes
and finite objects since they describe spinors on
pseudo-Riemannian manifolds as well as
finite geometries~\cite{Shaw-89}. Therefore, it
is not suprising that real Clifford algebras
can be used to define the  space
and time dimensions of an algebra  
representing (in a generalized sense)
a possibly noncommutative spacetime.
The present paper is a precise formulation
of this idea and the main ingredient of the
definition of a time dimension is the 
fundamental symmetry $\eta$ which allows for a
kind of Wick rotation of spacetime.

The fermionic and gauge Lagrangian
of particle physics models were given here
for Lorentzian spacetimes, but the derivation of
the full spectral action is still a difficult
open problem.

For applications to topological insulators,
it would be desirable to extend these results
to the case of odd-dimensional algebras. 

\section{Acknowledgments}
We are extremely grateful to Harald Upmeier for sending
us a copy of Elsner's Master's thesis.
We thank 
Shane Farnsworth, 
Johannes Kellendonk,
Guo Chuan Thiang,
Karen Elsner,
Franciscus Jozef Vanhecke,
Bertfried Fauser,
Jordan Fran{\c{c}}ois
for discussions
and correspondence.

\bibliographystyle{elsart-num}
\bibliography{qed}

\end{document}